# Enhanced and Spectrally Selective Near Infrared Photothermal Conversion in Plasmonic Nanohelices

*Juan A. Delgado-Notario[1], David López-Díaz[2], David McCloskey[3], José M. Caridad[1,4]\**

[1]Departamento de Física Aplicada, Universidad de Salamanca, 37008 Salamanca, Spain

[2]Departamento de Química Física, Universidad de Salamanca, 37008 Salamanca, Spain

[3]School of Physics, CRANN and AMBER, Trinity College Dublin, The University of Dublin,

[4]Unidad de Excelencia en Luz y Materia Estructurada (LUMES), Universidad de Salamanca, 37008 Salamanca, Spain

e-mail: jose.caridad@usal.es

We study the photothermal conversion in plasmonic nanohelices, unveiling how helical nanostructures made from metals with a notable interband activity -such as cobalt (Co) and nickel (Ni)- exhibit a remarkable temperature rise $\Delta T$ up to ~1000 K under illumination.

Such outstanding $\Delta T$ values exclusively occur at wavelengths close to their localised plasmon resonances ($\Delta T$ is significantly lower off resonance), and therefore the photothermal conversion of these nanoparticles is spectrally selective.

The exceptional and spectrally selective temperature rise is demonstrated at near infrared wavelengths, which prompts the use of Co and Ni helical nanoparticles in a wide range of photothermal applications including solar energy conversion, seawater desalination, catalysis, or nanomedicine.

## 1. Introduction

In recent years, plasmonic photothermal conversion has emerged as an attractive energy transformation technique with potential use in a wide range of applications.[1-4] The effect takes advantage of localised surface plasmon resonances (LSPRs) occurring in different nanostructured substrates across the ultraviolet (UV), visible (Vis) and near-infrared (NIR) region of the electromagnetic spectrum. Such resonances allow the effective absorption of electromagnetic energy in the nanomaterials, which is ultimately converted into heat via nonradiative plasmon decay.[3-5]

Despite being a well understood effect, the realization of functional plasmonic nanostructures with efficient photothermal conversion is not a straightforward task, especially when operating at NIR frequencies. In particular, an effective transformation of electromagnetic energy into heat is commonly achieved in nanoparticles with small sizes (tens of nm in all spatial directions).[4,5] Nonetheless, these nanostructures commonly exhibit LSPRs at UV and Vis frequencies, and thus their optimal photothermal conversion is limited to such energy ranges. Larger nanoparticles (exceeding hundreds of nm in a spatial direction) do present LSPR at NIR frequencies, however, they predominantly exhibit a radiative plasmon decay rather than an efficient light-to-heat conversion.[4,5] Strategies to design plasmonic nanoparticles with a large photothermal conversion at NIR frequencies should therefore not only consider the particle morphology (size or shape), but also additional and relevant parameters such as the material composition.[6]

In this study, we show an outstanding photothermal conversion efficiencies at NIR frequencies in weakly-coupled nanohelical particles made from transition metals such as cobalt (Co) and nickel (Ni). Such enhanced photothermal response can be ascribed to the joint contribution of three distinct factors. First, the unique (helical) shape of the nanoparticles allows to easily position LSPRs of the nanostructured substrates at the desired frequencies (NIR in this study).[7,8] Second, the weak interparticle plasmon coupling existing in arrays of nanohelices, separated at distances $\delta$ comparable to the diameter of the wire forming the nanohelix $d_w$ ($\delta \sim d_w$), is able to increase the overall light-harvesting efficiency of the nanoparticles.[9,10] Third, we demonstrate that, by using metals with relevant interband transitions up to IR frequencies (Co and Ni),[11,12] the photothermal conversion efficiency of metal nanohelices is notably enhanced and an outstanding temperature rise $\Delta T$ is achieved in these nanostructures. Indeed, $\Delta T$ reached when illuminating Co and Ni nanohelices at 785nm (wavelength close to the position of the LSPRs of these nanoparticles) is estimated to be up to ~1000 K and ~650K, respectively. This is in clear contrast with the notably lower values of $\Delta T$ measured in metal nanohelices with a similar shape but made of more conventional plasmonic materials free of interband transitions at IR frequencies such as silver (Ag). Moreover, we demonstrate that the photothermal conversion in Co and Ni nanohelices is spectrally selective. This is reflected in the considerably smaller temperature rise measured when illuminating these samples at wavelengths away from the LSPR (*ΔT* values for Co and Ni nanohelices measured at 633 nm are only ~50K and ~60K, respectively).

Overall, the temperature rise attained by these nanostructured substrates is substantially larger than those reported in other nanomaterials, including plasmonic nanoparticles consisting of diverse combinations of morphology and constituting materials,[1,2,13] as well as other nanostructures exhibiting state-of-the-art photothermal efficiency based on different (non-plasmonic) light-to-heat conversion mechanisms.[14] As such, our work reveals nanohelices made from metals with large interband activity as promising nanostructures to be used in a wide range of applications including biomedicine, catalysis or areas related to solar energy harvesting.[1-4]

## 2. Results

### 2.1. Photothermal conversion in Co and Ni nanohelices

We estimate the temperature of arrays of Co and Ni nanohelices upon illumination by measuring the surface-enhanced Raman scattering (SERS) spectra of graphene as common probing molecule placed on top of metal nanoparticles (methods).[1,15,16] In nanomaterials without or with a small SERS enhancement factor EF (~1, case of Co and Ni nanohelices)[16] the Stokes/anti-Stokes intensity ratio ($\frac{I_S}{I_{aS}}$) of the Raman spectrum of probing molecules depends primarily on the population of the vibrational levels involved in the inelastic scattering processes, which is a function of the local temperature of the sample $T$:[1,17,18]

$$\frac{I_S}{I_{aS}} = e^{\frac{\hbar\omega}{k_B T}} \quad \text{(Equation 1),}$$

where $k_B$ is the Boltzmann constant and $\hbar\omega$ is the energy of the Raman mode in consideration. The two most prominent features of the Raman spectra of graphene are the so-called G and 2D modes which lie around ~ 1580cm$^{-1}$ and ~ 2650cm$^{-1}$, respectively.[19] We use the G mode (energy $\hbar\omega_G \approx 195$ meV) to estimate the local temperature of the graphene sample $T$ and therefore record both Stokes and anti-Stokes Raman lines at ~ 1580cm$^{-1}$ and ~ -1580cm$^{-1}$, respectively. Prior to study the spectra of graphene placed on the fabricated nanohelices, we measure (methods) the Raman spectrum of graphene flakes on two reference substrates:[15,16,19] a Si substrate with 300nm SiO$_2$ on top and polydimethylsiloxane, PDMS. The measured ratio of the Raman peaks of graphene on both Si/SiO$_2$ and PDMS (see measurements in Supporting Note 1) is $\frac{I_S}{I_{aS}} \sim$ 45-47, which corresponds to a sample temperature $T$~600 K (value which is consistent with those measured in literature for graphene-based materials under similar measurement conditions)[1] . Moreover, we have also measured the Stokes to anti-Stokes intensity ratio of the Raman G band of graphene on a 50 nm Co thin film (see measurements in Supporting Note 1), obtaining similar $\frac{I_S}{I_{aS}}$ between 40-50. The fact that the estimated local temperature values are $T\sim 600$ K for graphene on all these non-structured substrates (Si/SiO$_2$, PDMS and Co film, three materials with different thermal conductivities) indicates that, regardless of the substrate, in-plane heat transfer predominantly occurs in these graphene samples. Such behaviour is a consequence of the high thermal conductivity of this two-dimensional material.

Figure 1a shows a schematic of the helical arrays used in the study (left panel), together with scanning electron microscopy images of the fabricated Co and Ni nanohelices (middle and right panels, respectively). The nanohelices exhibit a relatively uniform pitch $p$ and diameter $D$ ($p \sim$ 140 nm and $D \sim$ 95 nm, respectively) and have a similar number of turns $N \sim 2.5$. The plasmonic behaviour of metallic nanohelices is dominated by a longitudinal LSPR approximately placed at a wavelength $\lambda_p$ equal to the total length of the wire composing the nanohelix,[7,8]

$$\lambda_p \sim N\sqrt{(\pi D)^2 + p^2} \quad \text{(Equation 2)},$$

and thus the longitudinal LSPR in the samples presented in Figures 1b,c is located at NIR frequencies (~824 nm). Reflectance measurements have been additionally carried out in our samples to experimentally verify the position of their longitudinal LSPR (see Supporting Note 2).

As reported in literature,[7,9,10,16] at wavelengths close to their longitudinal LSPR, the near-field distribution of multiple-turn nanohelices consists of periodic dipoles (with an effective length of half a pitch, $p/2$) along the length of the nanostructures. The interaction between these dipoles induces a notable light absorption in the individual nanohelices at wavelengths close to the LSPR.[7,9] Moreover, the nanohelices presented in this study are weakly coupled since they are arranged in an array configuration with a separation distance between nanoparticles ($\delta\sim$ 75 nm) which is comparable with the diameter of the wire forming the nanohelices $d_w$. Metal nanohelices with these aspect ratios $\delta/d_w \sim 1$ provide an increased light-harvesting efficiency w.r.t. the individual nanoparticle.[9,16] Such absorbed electromagnetic energy can be ultimately re-radiated or converted into heat depending on, for instance, the interband activity of the metals constituting the nanostructures.[7,9,16]

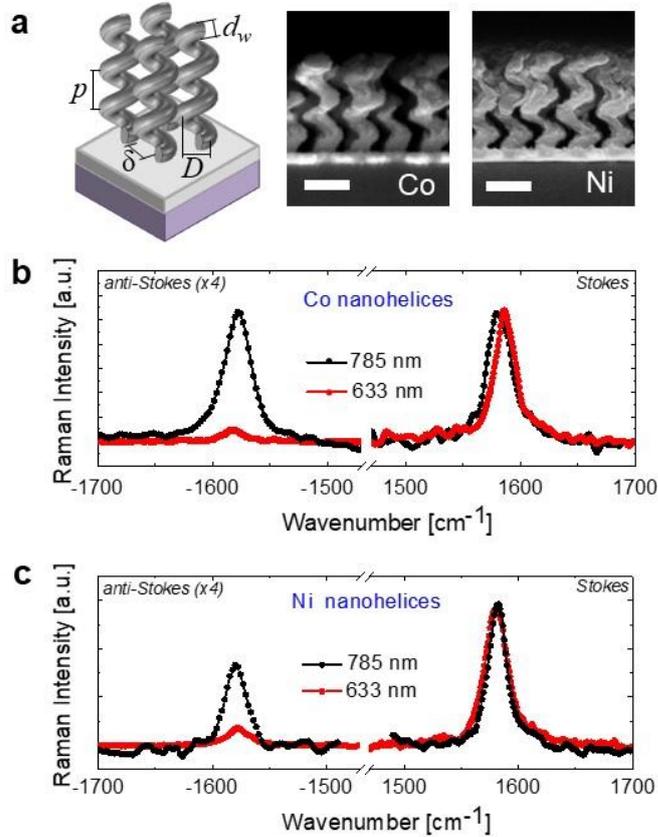

Figure 1. (a) Left panel: schematic showing the main parameters of the helical array including the helical diameter $D$, pitch $p$, wire diameter $d_w$ and separation between helices δ. Middle and right panels show scanning electron micrograph images of the arrays of Co and Ni nanohelices, respectively. Scale bars in both insets are 150 nm. (b) anti-Stokes and Stokes Raman G-line of graphene on Co nanohelices and for 633 nm (red) and 785 nm (black) laser wavelengths. (c) anti-Stokes and Stokes Raman G-line of graphene on Ni nanohelices and for 633 nm (red) and 785 nm (black) laser wavelengths.

Figures 1b and 1c depict the measured Stokes and anti-Stokes G mode Raman lines of graphene placed on arrays of Co and Ni nanohelices, respectively, at two different laser wavelengths (785 nm and 633 nm). At a first glance, both arrays of Co and Ni nanohelices show much smaller $\frac{I_S}{I_{aS}}$ ratios at 785 nm than at 633 nm laser wavelengths. This fact stems from an enhanced light-harvesting occurring in arrays of nanohelices at wavelengths close to the position of their LSPR,[7] (~824 nm) and it makes evident the spectrally selective photothermal conversion provided by these nanostructures. In particular, the estimated temperatures of graphene on nanohelices at 633 nm wavelengths (i.e. away from the position of the LSPR) are $T$~620 K ($\frac{I_S}{I_{aS}}$

~ 38) and $T$~660K ($\frac{I_S}{I_{aS}}$ ~ 30) for Co and Ni nanohelices, respectively. Therefore, helical nanostructures induce a minimal temperature rise $\Delta T$ of 20 K (Co nanohelices) and 60 K (Ni nanohelices) at 633 nm. Instead, the estimated temperatures for graphene on Co and Ni nanohelices at 785 nm wavelength are notably larger, $T$~1600 K ($\frac{I_S}{I_{aS}}$ ~ 4) and $T$~1250 K ($\frac{I_S}{I_{aS}}$ ~ 6), respectively. The temperature rise $\Delta T$ induced by Co and Ni nanohelices at 785 nm is then 1000 K and 650 K, respectively. These values are notably larger than those obtained at 633 nm, despite the fact that the laser irradiance $I$ at 785 nm (0.12 mW/µm²) is an order of magnitude smaller than the one at 633 nm (6.6 mW/µm²), see methods for further measurement details. We remark at this point that the estimated $\Delta T$ values for Ni and Co nanohelices are accurate: the temperature rise of the nanoparticles is the primary factor contributing to a decrease in the measured $\frac{I_S}{I_{aS}}$ ratio in plasmonic nanostructures with SERS EF ~1,[17,18] which is the case of Ni and Co nanohelices.[16] Moreover, we highlight that *i)* the temperature rise $\Delta T$ increases approximately linearly with the irradiance of the laser *I* (see Suppporting Note 3) and *ii)* the temperature rise obtained for both Co and Ni nanohelices at 785 nm and an irradiance 0.12 mW/µm² is large, roughly half of the melting point temperatures of the bulk materials (1768 K and 1738 K for Co and Ni, respectively).[18] In fact, following these two observations, one can also understand that a disintegration of these nanohelices occurs below the laser spot when increasing the irradiance of the 785 nm laser to 0.3 mW/µm² (see Supporting Note 4).

Next, we show that the combined effect of helical nanostructures made from metals with relevant interband transitions are the primary cause of the efficient light-to-heat conversion in Ni and Co nanohelices. To do so, we estimate the temperature rise of an array of nanohelices made from a different metal, Ag. Contrary to Co and Ni, Ag is a conventional plasmonic material with no major interband activity at wavelengths above ~400 nm,[20] and the plasmonic decay in Ag nanohelices is predominantly radiative through scattering.[7,9,16] In this sense, it is expected that the local temperature reached in arrays of Ag nanohelices is notably lower than the one attained in arrays of Co and Ni nanohelices. Figure 2 shows the measured Stokes and anti-Stokes Raman lines of graphene placed on the array of Ag nanohelices. Ratios $\frac{I_S}{I_{aS}}$ ~30 at 633 nm wavelength and $\frac{I_S}{I_{aS}}$ ~15 at 785 nm wavelength are extracted for these samples, which correspond to estimated temperatures ~ 660 K and ~830 K and a temperature rise of $\Delta T$ 60K and 230 K, respectively. We further highlight that these estimated temperature values are an upper limit for the case of Ag nanohelices: factors other than a temperature rise of the nanohelices contribute to a decrease in the measured $\frac{I_S}{I_{aS}}$ ratio[17,18] in plasmonic nanostructures with SERS EF >>1 such as Ag nanohelices[9,16] (see further information in Supporting Note 5).

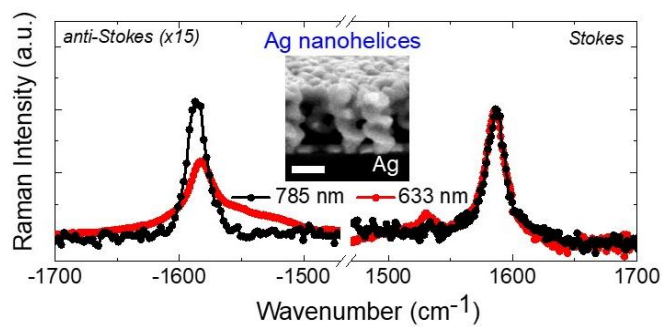

Figure 2. Anti-Stokes and Stokes Raman G-line of graphene on Ag nanohelices for 633 nm (red) and 785 nm (black) laser wavelengths. Inset shows a scanning electron micrograph image of the array of Ag nanohelices. The scale bar in the inset is 150 nm.

## 2.2. Performance of Co and Ni nanohelices as efficient photothermal conversion devices for applications

The remarkable temperature rise $\Delta T$ achieved in Co and Ni nanohelices at 785 nm (~1000 K and ~650 K at a laser irradiance 0.12 mW/µm$^2$, respectively) promotes these nanostructured substrates as effective photothermal conversion devices at NIR frequencies. These nanomaterials are particularly interesting for high temperature photothermal applications, systems where typical working temperatures >500 K are required.[3-5] In fact, $\Delta T$ values reached in Co and Ni helical nanoparticles are larger than state-of-the-art nanostructures reported in literature.[14]

A detailed benchmarking between the here reported Co and Ni nanohelices and other photothermal nanomaterials is provided in Figure 3. For a proper comparison, we only consider here substrates measured in similar conditions than our nanohelices (i.e. using monochromatic radiation and atmospheric conditions), excluding thus nanomaterials measured in solution or studies where polychromatic radiation was used.[21,22] In particular, Figure 3 depicts the temperature rise achieved in the different photothermal nanosystems normalized w.r.t. the irradiance $I$ of the incident visible or NIR radiation $\overline{\Delta T} = \Delta T / I$ (raw data for composing this image is shown in Supporting Note 6). The figure itself is separated into three regions depending on the melting point of the bulk materials constituting the nanostructures $T_m$: below 750K (left hand side of the panel, green region), between 750K and 1500K (middle part of the panel, orange region) and above 1500K (right hand side of the panel, red region). In turn, these three regions classify photothermal conversion devices/substrates according to their ability to operate at low, medium or high working temperatures, respectively.

Intriguingly, Co and Ni nanohelices at 785 nm show the largest $\overline{\Delta T}$ of the substrates able to work at high temperatures (red region), surpassing state-of-the art substrates such as silicon nanospheres by an order of magnitude.[14] Nanostructured substrates based on polymers show the largest $\overline{\Delta T}$ among all substrates,[23-25] however, they can only be utilized at much lower temperatures due to their low melting point (i.e. the temperature rise $\Delta T$ reached in these substates is also low). Figure 3 also illustrates the fact that arrays of weakly coupled Co and Ni nanohelices excel at spectrally selective photothermal conversion i.e. at efficiently converting electromagnetic radiation into heat only at specific wavelengths: the normalized temperature rise exhibited by the nanostructures at 633 nm is three orders of magnitude smaller than the temperature rise at 785 nm (close to the longitudinal LSPR).

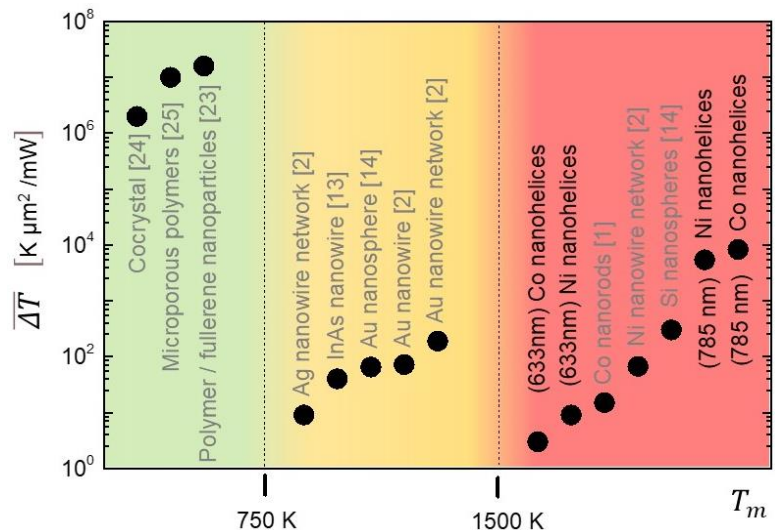

Figure 3. Temperature rise normalized w.r.t. the irradiance of the incident radiation for different photothermal systems: nanohelices reported in this work (black letters) and in literature (grey letters). The graph is separated in three regions, depending on the melting point of the constituting bulk materials: below 750K (green part of the panel, left hand side), between 750K and 1500K (orange part of the panel, in the middle) and above 1500K (red part of the panel on the right hand side).

Furthermore, we note that metals such as Ni and Co have lower surface diffusion than more conventional plasmonic materials such as Ag, aluminum (Al) or gold (Au).[26,27] This fact minimizes the reshaping of Ni and Co nanohelices at temperatures close to their melting threshold and thus aids in the stability of these novel photothermal substrates.

In terms of applications, not only the fact of reaching a large temperature rise $\Delta T$ at specific wavelenghts in the here reported Co and Ni nanohelices is important. We also remark the possibility of controlling such temperature rise in the surroundings of these nanostructured materials depending on the irradiance of the incoming light (see Supporting Note 3). Finally, we accentuate the fact that the plasmonic nanohelices disintegrate even at modest laser irradiances (above 0.30 mW/µm$^2$ , see Supporting Note 4), which makes these nanomaterials readily to be used in laser-induced explosion applications.[28]

## 3. Conclusion

In conclusion, by using micro-Raman spectroscopy, we have demonstrated Co and Ni nanohelices to be efficient photothermal conversion devices at IR wavelengths. The measured temperature rise in these nanostructures when using a relatively low laser irradiance (0.12 mW/µm$^2$) reaches 1000K, value which is larger than the one reported in state-of-the-art photothermal nanomaterials able to operate at high temperatures.[14] Such efficient light to heat conversion is achieved at specific wavelengths close to the LSPR of the nanostructures and thus spectrally selective. The simple tuning of the LSPR of metallic nanohelices,[7] and the scalable fabrication of these nanomaterials,[29] prompt the use of these novel photothermal substrates in a wide range of applications including solar energy conversion, photothermal signal generation in radiometry and nanomedicine or laser-induced explosion.[1-5,28] Moreover, being chiral systems, nanohelices exhibit a different absorption of right-handed and left-handed circularly polarized light,[30] behaviour which can be potentially exploited to design unique and novel devices with chiral photothermal properties.

**Experimental Section/Methods**

*Fabrication of nanohelices*

Periodic arrays of helical nanostructures made from Ni, Co and Ag are engineered by oblique angle deposition, OAD.[7,29] The OAD set-up for this study is implemented in a general purpose evaporator by means of an in-house built retrofittable system which provides controlled substrate rotation. We use the OAD method on seeded substrates (150 nm separation) with a ~50nm Ag film on top to produce the three regular nanohelical arrays made from Co, Ni and Ag.[29] The oblique angle and rotation speed were fixed at ~ 88º and 0.3 rpm, respectively for all the samples and the deposition rate was 1.5 nm/s. Scanning electron microscopy (SEM) was utilized to examine the structure and morphology of the grown metallic nanohelices.

*Graphene transfer on nanohelices*

Polymethyl methacrylate (PMMA) is spun on top of Monolayer CVD graphene on Cu foil (from Graphenea). The Cu is etched in a 1M ammonium persulfate solution. Graphene supported with PMMA is then transferred on the helical nano-arrays. Finally, the PMMA layer is dissolved in acetone, and samples are rinsed in isopropanol and dried with $N_2$ gas.

*Raman measurements*

Raman spectra were measured with a micro-Raman set-up (Horiba Jobin-Ybon Labram HR) with ~2 cm$^{-1}$ spectral resolution, using a 100× objective with numerical aperture N.A. 1.25. The excitation wavelengths were 633 nm (laser spot ~ 1μm$^2$) and 785 nm (laser spot ~ 12 μm × 6μm). Unless otherwise stated, by using a filter which transmits 10% of the incoming light, the laser power on the samples was set in our measurements to 6.6 mW for both laser lines and thus their irradiance is ~6.6 mW/ μm$^2$ and ~0.12 mW/ μm$^2$ for 633 nm and 785 nm wavelength, respectively. The acquisition time for each spectrum was 2 minutes and, to ensure reproducibility and stability, we measured five Raman spectra for each array of nanohelices. When needed (i.e. to undertake measurements at different laser powers), additional filters transmitting 1%, 5% or 25% of the incoming light were used in the Raman spectrometer.

**Acknowledgements**

Authors thank the support from the Ministry of Science and Innovation (MCIN) and the Spanish State Research Agency (AEI) under the grant PID2021-128154NA-I00 funded by MCIN/AEI/10.13039/501100011033 and by "ERDF A way of making Europe". This work has been also supported by Junta de Castilla y León co-funded by FEDER under the Research Grant number SA103P23. J.A.D.-N. thanks the support from the Universidad de Salamanca for the María Zambrano postdoctoral grant funded by the Next Generation EU Funding for the Requalification of the Spanish University System 2021–23, Spanish Ministry of Universities. J.M.C. acknowledges financial support by the MCIN and AEI "Ramón y Cajal" program (RYC2019-028443-I) funded by MCIN/AEI/10.13039/501100011033 and by "ESF Investing in Your Future". J.M.C. also acknowledges financial of the European Research Council (ERC) under Starting grant ID 101039754, CHIROTRONICS, funded by the European Union. Views and opinions expressed are however those of the author(s) only and do not necessarily reflect those of the European Union or the European Research Council. Neither the European Union nor the granting authority can be held responsible for them. D. McC. acknowledges funding under Science Foundation Ireland's Frontiers for the Future program grant no. 19/FFP/6745.

# Supporting Information

# Enhanced and Spectrally Selective Near Infrared Photothermal Conversion in Plasmonic Nanohelices

*Juan A. Delgado-Notario[1], David López-Díaz[2], David McCloskey[3], José M. Caridad[1,4]\**

[1]Departamento de Física Aplicada, Universidad de Salamanca, 37008 Salamanca, Spain
[2]Departamento de Química Física, Universidad de Salamanca, 37008 Salamanca, Spain
[3]School of Physics, CRANN and AMBER, Trinity College Dublin, The University of Dublin,
[4]Unidad de Excelencia en Luz y Materia Estructurada (LUMES), Universidad de Salamanca, 37008 Salamanca, Spain

e-mail: jose.caridad@usal.es

**Supporting Note 1 – Raman of graphene on non-structured substrates (Si/SiO₂, PDMS and Co film)**

Figure S1 shows the measured anti-Stokes (left panel) and Stokes (right panel) G mode Raman lines of graphene placed on SiO$_2$/Si substrates. Moreover, the right panel also includes the characteristic 2D band of graphene, appearing at ~2630 cm$^{-1}$. The spectra is only measured at 633 nm, since the Raman signal of graphene on SiO$_2$/Si at this wavelength shows a broad fluorescent band of the SiO$_2$ at 785 nm, avoiding the observation of the Raman modes.[S1] By taking into account the interference-related enhancement of the Raman signal occuring in graphene on SiO$_2$ (a factor ~2 at 633nm),[S2] the measured $\frac{I_S}{I_{aS}}$ ratio is ~46.

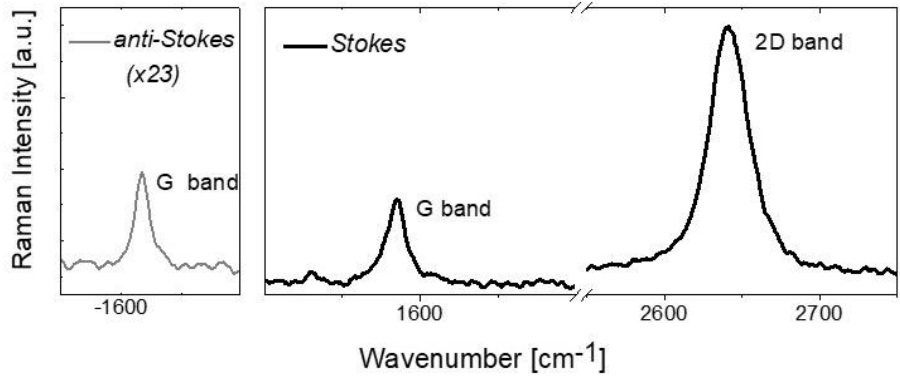

Figure S1. Raman spectra of graphene on a SiO$_2$/Si substrate. Anti-Stokes spectra close to the Raman G peak (left panel) and Stokes spectra close to the Raman G and 2D peaks (right panel) at an excitation wavelength of 633 nm. $\frac{I_S}{I_{aS}}$ ratio in these samples is ~46, taking into account a ×2 factor due to interference-related enhancement of the Raman signal of graphene on SiO$_2$/Si. [S2]

Figure S2 shows the measured anti-Stokes (left panel) and Stokes (right panel) G mode Raman lines of graphene placed on polydimethylsiloxane, PDMS. These Raman spectra are obtained at two different laser wavelengths (785 nm and 633 nm), and show $\frac{I_S}{I_{aS}}$ ratios between ~45 and ~47 for both laser lines.

$\frac{I_S}{I_{aS}}$ ratios are therefore similar in the two used reference substrates and correspond to an estimated temperature of ~600K. The fact that the estimated local temperature values of graphene on the two substrates Si/SiO$_2$ and PDMS is coincident $T$~600 K indicates that, regardless of the substrate, in-plane heat transfer primarily occurs in these samples. Such behaviour agrees well with related studies in literature (placing graphene on top of different substrates, including metals), [S3] and is a consequence of the high thermal conductivity of this two-dimensional material.

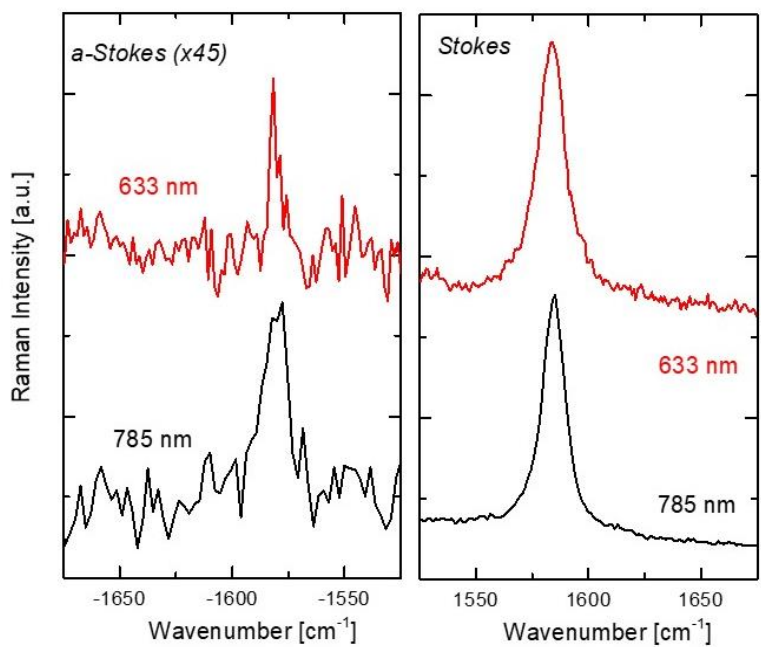

Figure S2. anti-Stokes and Stokes Raman G-line of graphene on PDMS for 633 nm (red) and 785 nm (black) laser wavelengths.

For completeness, Figure S3 depicts the measured anti-Stokes (left panel) and Stokes (right panel) G mode Raman lines of graphene placed on a 50 nm thick Co film. This film is relevant: it is made from the same type of metal than some of the studied arrays of nanohelices, but is absent of localised surface plasmon resonances existing in the nanohelices. Albeit the existing noise in the anti-Stokes measurements, Raman spectra obtained at the two different laser wavelengths (785 nm and 633 nm) also shows $\frac{I_S}{I_{aS}}$ ratios between ~40-50 for both laser lines (estimated temperatures between ~580K and ~610K). The fact that the estimated temperature of graphene on the Co film is *i)* similar at both wavelengths 633 nm and 785 nm; *ii)* similar to the estimated local temperature in Co and Ni nanohelices at 633 nm (~620 K and ~660K, respectively); but *iii)* very distinct to the estimated temperature of Co and Ni nanohelices at 785 nm (~1600K and ~1250K, respectively) highlights the importance of the longitudinal LSPR and the nonradiative plasmon decay as key mechanisms to achieve an efficient and spectrally selective photothermal conversion in helical nanostructures.

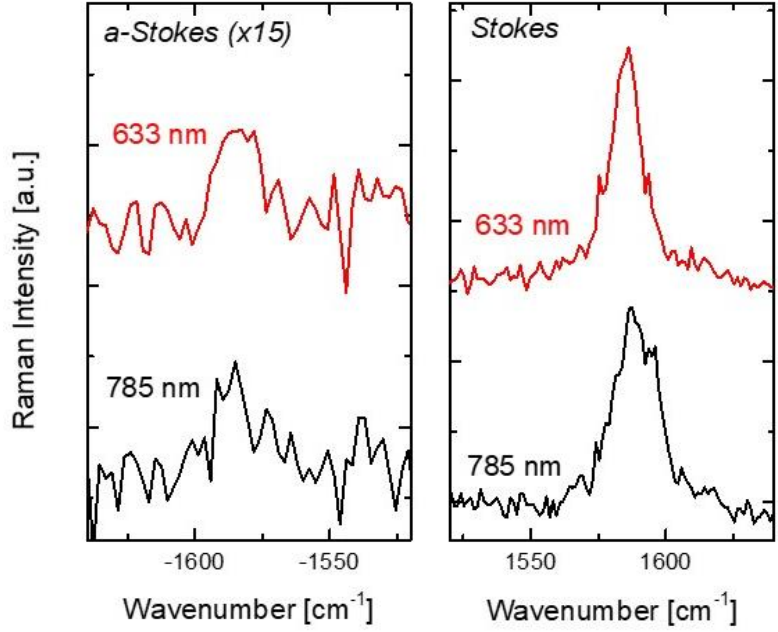

Figure S3. anti-Stokes and Stokes Raman G-line of graphene on a 50 nm Co film for 633 nm (red) and 785 nm (black) laser wavelengths.

Finally, we note that graphene was stable during our experiments (on all substrates Si/SiO$_2$, PDMS, Co film and the nanohelices). We verify this by ensuring that the D band remains stable during all recorded Raman spectra. Such band appears at ~1300 cm$^{-1}$ and its stability indicates that no new defects are created in the monolayer during the undertaken measurements. This is also consistent with previous works in literature reporting the ablation of graphene at visible/NIR wavelengths using lasers with fluences > 30 mJ/cm$^2$ (pulse duration 10 -100 ps).[S4] The latter corresponds to irradiances > 3 ×10$^3$ mW/µm$^2$ which are three orders of magnitude larger than laser irradiances used in our experiments.

**Supporting Note 2 - Reflectance measurements on Ag, Co and Ni nanohelices**

Far-field reflectance measurements have been undertaken in our samples at normal incidence to experimentally verify the position of the longitudinal LSPR. In brief, the reflectance of helical nanostructures is composed by two main minima.[S5] The one at higher wavelength (~2000 nm) indicates the effective medium threshold. The second minima appears at shorter wavelengths and indicates the position of the longitudinal LSPR wavelength $\lambda_p$.[S5] Figure S4 depicts the reflectance minima for the fabricated Ag, Co and Ni nanohelices placed between 800 and 830 nm. These values are in good agreement with the scaling law shown in previous far-field reports, [S5] (Equation 2 of the main text): $\lambda_p \sim lN$ where $l = \sqrt{(\pi D)^2 + p^2}$ is the length of one turn and $N$ the number of turns of the individual nanohelix.

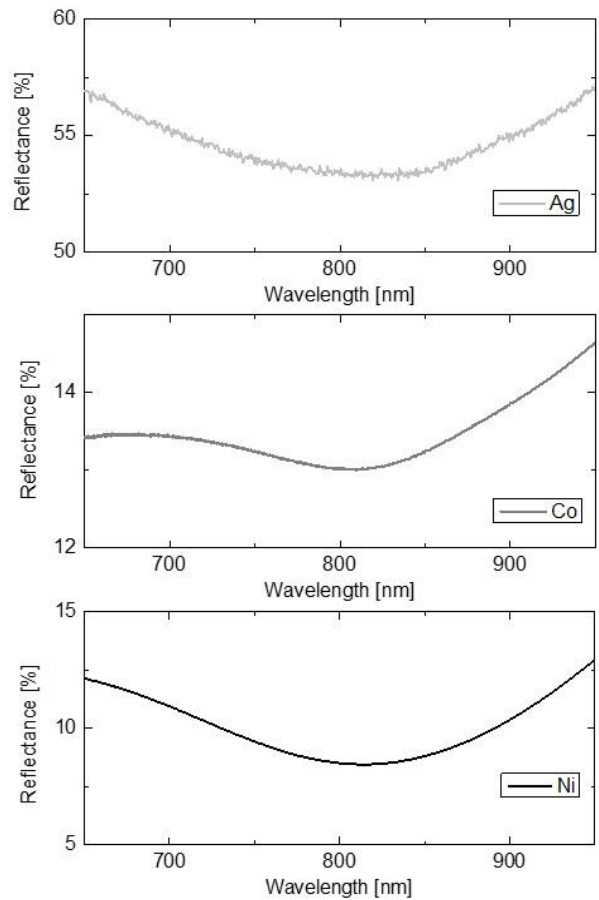

Figure S4. Reflectance measurements of Ag (top panel), Co (middle panel) and Ni (bottom panel) nanohelices used in the study.

**Supporting Note 3 – Dependence of the temperature rise $\Delta T$ with laser irradiance $I$**

Figure S5 shows the dependence of the laser irradiance $I$ with the temperature rise $\Delta T$ at 785 nm wavelength. In particular, three irradiances are used: 0.12, 0.6 and 0.012 mW/µm² which correspond to 10%, 5% and 1% of the total laser power offered by our Raman tool, respectively (see methods). To a first approximation, $\Delta T$ increases linearly with $I$ in both Co and Ni nanohelices.

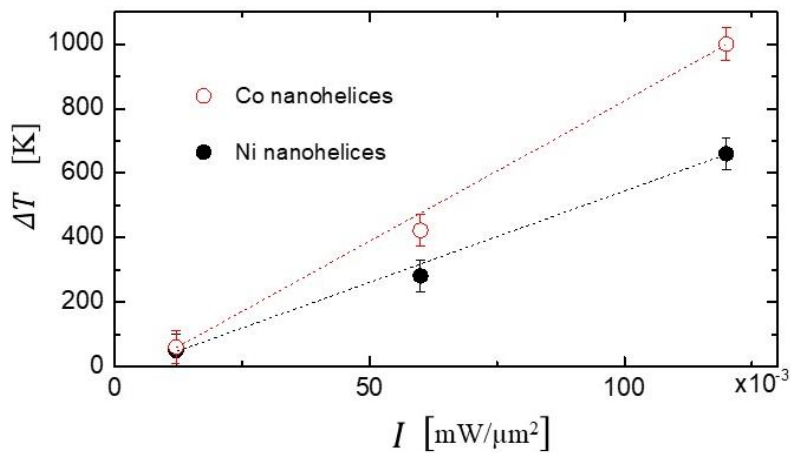

Figure S5. Dependence of the temperature rise $\Delta T$ w.r.t. the laser irradiance $I$ for both Co (red) and Ni (black) nanohelices. The laser wavelength is 785 nm. Error bars in the figure are given by the noise observed in the anti-Stokes band.

**Supporting Note 4 – Melting of nanohelices at a higher laser irradiances**

Figure S6a shows a top view of Ni nanohelices with graphene on top. The matt-black appearance indicates strong visible-light absorption taking place in these nanostructures.[S5,S6] The graphene film (right hand side of the panel) can be clearly distinguished w.r.t. areas of the substrate not covered by this material (left hand side of the image). Figure S6b shows a top view of Ni nanohelices with graphene on top, where the nanostructures have disintegrated below the laser spot upon illumination at 785 nm and an irradiance of 0.30 mW/µm$^2$ (corresponding to 25% of the total laser power in our tool). This fact agrees well with the approximately linear increase of the temperature rise $\Delta T$ observed in these nanostructures w.r.t. the laser irradiance $I$ (Figure S5) and can be attributed to a local temperature surpassing the melting point of bulk Ni (1726 K) at the aforementioned irradiance 0.30 mW/µm$^2$.

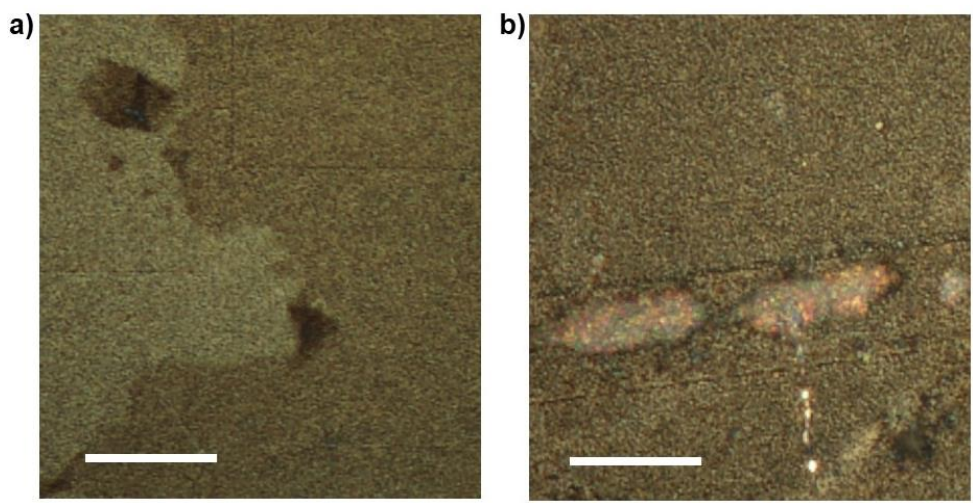

Figure S6. a) Optical image of Ni nanohelices. Graphene on top of nanostructures (right hand side of the figure) is visible by the different contrast present in the image. b) Optical image of the array of Ni nanohelices after the sample has been illuminated with a 785 nm laser line (laser spot ~ 12 µm × 6 µm) at an irradiance 0.3 mW/µm$^2$. Two ellipsoidal areas where nanohelices have been disintegrated can be observed. Scale bars in both panels are 10µm.

**Supporting Note 5 – Accuracy of temperature estimation for Ag nanohelices under illumination**

The measured $\frac{I_S}{I_{aS}}$ ratio of the Raman G peak of graphene on Ag nanohelices (Figure 2, main text) was found to be ~30 and ~15 for 633 nm and 785 nm excitation wavelengths, respectively. If such ratios are only ascribed to a temperature rise up of the nanostructures, these values would lead to a temperature estimation $\Delta T$ of ~650 K (at 633 nm) and ~800 K (at 785nm). However, such estimated temperature values for Ag nanohelices only represent an upper limit since the enhancement factor of these nanostructures is larger than unity[S5-S9]. In cases where EF>>1, the measured $\frac{I_S}{I_{aS}}$ ratio will not only depend on the temperature of the nanostructures but also on vibrational pumping and resonance effects. These two additional effects will increase the efficiency of anti-Stokes processes.[S7,S8] Since all of these factors will contribute to decrease $\frac{I_S}{I_{aS}}$, the temperature estimation for the case of Ag nanohelices therefore is an upper limit.

We remark that this case does not apply for Co or Ni nanohelices, nanostructures with no near-field enhancement EF~1.[S9] Therefore the temperature estimation in Co and Ni nanohelices is accurate using $\frac{I_S}{I_{aS}}$.[S6,S7]

**Supporting Note 6 – Comparison of Co and Ni nanohelices with state-of-the-art photothermal nanostructures**

Table 1. Table reporting the working temperature $T$, temperature rise $\Delta T$, laser irradiance and normalized temperature rise $\overline{\Delta T}$ in different photothermal devices.

| Nanostructured material | $T$ [K] | $\Delta T$ [K] | Laser Power [mW] | Laser spot [$\mu m^2$] | Laser Irradiance [mW/$\mu m^2$] | $\overline{\Delta T}$ Norm temperature increase [K $\mu m^2$ /mW] |
|---|---|---|---|---|---|---|
| Co nanohelices (785 nm) | 1600 | 1000 | 6.6 | ~55 | 0.12 | $8.3 \times 10^3$ |
| Co nanohelices (633nm) | 620 | 20 | 6.6 | ~1 | 6.6 | 3 |
| Ni nanohelices (785 nm) | 1250 | 650 | 6.6 | ~55 | 0.12 | $5.4 \times 10^3$ |
| Ni nanohelices (633nm) | 660 | 60 | 6.6 | ~1 | 6.6 | 9 |
| Co nanorods Ref [1], main text | 699 | 91 | 6 | ~1 | 6 | 15 |
| Single Au nanowire Ref [2], main text | 500 | 203 | - | - | 2.82 | 72 |
| Au nanowire network Ref [2], main text | 420 | 120 | - | - | 0.63 | 190 |
| Ni nanowire network Ref [2], main text | 410 | 110 | - | - | 1.65 | 67 |
| InAs nanowires Ref [13], main text | 360 | 60 | 0.5 | 0.34 | 1.47 | 40 |
| Si nanospheres Ref [14], main text | 900 | 600 | - | - | 2 | 300 |
| Au nanospheres Ref [14], main text | 430 | 130 | - | - | 2 | 65 |
| Cocrystal Ref [24], main text | 70 | 45 | - | - | $2.3 \times 10^{-6}$ | $1.9 \times 10^6$ |

| Ag nanowire network Ref [2], main text | 325 | 25 | - | - | 2.82 | 8.8 |
| --- | --- | --- | --- | --- | --- | --- |
| Microporous polymers Ref [25], main text | 130 | 100 | - | - | $1\times10^{-5}$ | $1\times10^{7}$ |
| Polymer/fullerene nanoparticles Ref [26], main text | 27 | 50 | - | - | $3.2\times10^{-6}$ | $1.6\times10^{7}$ |